\documentclass[final,5p,times,twocolumn]{elsarticle}
\usepackage{amsmath,graphicx,float,hyperref,colortbl,upgreek,comment}

\begin{document}

\begin{frontmatter}

\title{Charm quark production in heavy-ion collisions as a signature of pre-equilibrium}

\address[irfu]{Universit\'e Paris-Saclay, Centre d’Etudes de Saclay (CEA), IRFU, IRFU, D\'epartement de Physique Nucl\'eaire (DPhN), Saclay, France}
\address[bielefeld]{Fakult\"at f\"ur Physik, Universit\"at Bielefeld, D-33615 Bielefeld, Germany}
\address[CERN]{CERN, EP Department, CH-1211 Geneva, Switzerland}

\author[CERN]{Maurice Coquet}
%\email{maurice.louis.coquet@cern.ch}

\author[irfu]{Thomas Faure}
%\email{Thomas.faure@ecolecentrale.fr}

\author[bielefeld]{Sören Schlichting}
% \email{sschlichting@physik.uni-bielefeld.de}

\author[irfu,bielefeld]{Mika Spier}
%\email{mspier@physik.uni-bielefeld.de}

\author[irfu]{Michael Winn}
% \email{michael.winn@cern.ch}

\begin{abstract}

The relative abundances and kinematic distributions of hadrons containing (anti)charm quarks are key observables for deconfinement, heavy-quark diffusion and hadronization in heavy-ion collisions. 
The production of (anti)charm  quarks is commonly associated to the initial hard scatterings in hadronic collisions. Based on previous studies on dilepton production, we evaluate the (anti)charm quark production from the pre-equilibrium phase. A non-negligible contribution to the overall charm quark production is found albeit large theoretical uncertainties are limiting factors. We conclude that precise total charm production measurements combined with progress on charm production calculations from the initial hard scatterings can be used to infer information on the pre-equilibrium stage. % In this contribution, we discuss the theoretical uncertainties, the implications of existing and future measurements  on the pre-equilibrium and the complementarity with respect to the dilepton production.

\end{abstract}

%\maketitle

\end{frontmatter}

\section{Introduction}
In ultrarelativistic heavy-ion collisions, a quark-gluon plasma (QGP) is formed which gradually thermalizes as it expands into the vacuum~\cite{Busza:2018rrf}. It is commonly assumed that this system approaches local thermal equilibrium through a pre-equilibrium phase during which the longitudinal pressure is smaller than the transverse pressure~\cite{Schlichting:2019abc,Berges:2020fwq}. 
This pressure anisotropy is generated by the rapid longitudinal expansion of the fluid at early times~\cite{Bjorken:1982qr,Ollitrault:2007du}. 
In addition, the system is initially gluon dominated~\cite{Lappi:2006fp}, implying that the pre-equilibrium phase is chemically out of equilibrium. % check if better references
These non-equilibrium features can be probed by electromagnetic radiation emitted throughout the evolution of the medium. In particular, dilepton production in the invariant mass range from 2 to 5 GeV/$c^2$, is sensitive to the pre-equilibrium phase via the production rate~\cite{Churchill:2020uvk, Coquet:2021lca,Coquet:2021gms,Plaschke2025, Wu:2024pba}, while the angular distribution of dileptons can be used to directly probe the pressure anisotropy~\cite{Coquet:2023wjk}.

In heavy-ion collisions, charm quarks are produced predominantly in initial hard scatterings, on time scales much shorter than those associated with the medium evolution, and the charm-quark number is expected to be conserved after the initial stages of the collision~\cite{Andronic:2006ky}. At hadronization, these quarks can form hadrons containing one charm quark or anti-charm quark (open charm hadrons), or charm-anti-charm bound states (charmonium).

On one hand, open charm hadrons represent an opportunity to trace individual color charge carriers propagating in the medium, and in particular can be used to constrain the diffusion of massive quarks in the QGP~\cite{Rapp:2018qla,Apolinario:2022vzg}. On the other hand, charmonia provide one of the most direct probes of deconfinement in heavy-ion collisions~\cite{Apolinario:2022vzg,Andronic:2025jbp}. In this context, the total charm production cross section in nucleus-nucleus collisions is of prime importance, as the total charm yield constitutes the natural normalization for charmonium production. Comparing the fraction of charm quarks that hadronize into charmonium or open-charm hadrons in heavy-ion collisions with the corresponding fractions measured in a reference system then provides a direct experimental access to medium-induced suppression or enhancement of charmonium production. At present, however, the total charm-production cross section remains one of the leading uncertainties in the interpretation of quarkonium measurements~\cite{Apolinario:2022vzg,Andronic:2025jbp}.
For these reasons, precise production measurements of all charm-hadron species are an important goal for the ongoing ALICE 2 and LHCb U1 programs as well as the future ALICE 3~\cite{Adamova:2019vkf,ALICE-PUBLIC-2025-005}, CMS and LHCb U2~\cite{LHCb:2018roe,Vagnoni:2025vmf} projects.
%CMS ref to be found, ATLAS to be checked if mentioned for ITK upgrade
%In general, the hadronisation of hadrons containing charm quarks is one of the limitation factors of the extraction of the diffusion coefficients. %LHCb Upgrade 2~\cite{LHCb:2018roe} and ALICE 3~\cite{Adamova:2019vkf}
%The total charm-quark production cross section remains one of the leading uncertainties in the interpretation of current quarkonium production measurements~\cite{Apolinario:2022vzg,Andronic:2025jbp}.

%It is commonly assumed that charm quark production originates from the initial hard scatterings of the collision. %check if better refeferences for this very point.
Although thermal charm production has been discussed in the literature and sizeable contributions have been found in explicit calculations~\cite{Zhang:2007yoa,Uphoff:2010sh,Zhou:2016wbo,Song:2024hvv}, % to be checked, if all calculations really THERMAL
charm production has not been discussed in the context of the pre-equilibrium phase. In this phase, the energy densities are highest, and hence the production of charm quarks can be expected to be particularly significant. This naive expectation is supported by the observation that charm quark production occurs at the same scales as intermediate mass dilepton production, where the pre-equilibrium phase plays an important role. %However, the out-of-equilibrium character of the pre-equilibrium .
%In the past, thermal charm production has 

We therefore propose that precise total charm production can be used to constrain the pre-equilibrium phase, by comparing measurements in nucleus-nucleus collisions with estimates from initial hard scatterings based on proton-proton data and nuclear parton densities.
We show that the charm production is complementary to dilepton production since it is also sensitive to the gluon abundance in the pre-equilibrium stage. To this end, we employ the same theoretical set-up as used for dilepton production~\cite{Coquet:2021lca, Coquet:2021gms,Plaschke2025}. We also discuss current theoretical limitations related to the extrapolation of the hard scattering production to nucleus-nucleus collisions and to the current state-of-the-art theory calculations. Finally, we outline how these limitations can be addressed in the future.

\section{Pre-equilibrium Charm Production}
We calculate the pre-equilibrium charm-anti-charm pair production to leading order in perturbation theory and employ an effective kinetic theory of QCD \cite{AMY2003} to describe the pre-equilibrium dynamics.  We denote by  $f_{g/q/\bar{q}} (\mathbf{p}_{1/2},x)$ the phase-space  distributions of gluons, quarks and anti-quarks, where ${\bf p}_1$ and ${\bf p}_2$ are the momenta of incoming particles, and $x$ the space-time coordinate at which the scattering occurs. The production rate of a charm-anti-charm-quark pair is
\begin{align}
    \nonumber\frac{dN^{c\bar{c}}}{d^4x d^4Q} &= \int \frac{d^3 p_1}{(2\pi)^3}\frac{d^3 p_2}{(2\pi )^3}\sum_f f_{q_f}(\mathbf{p_1}) f_{\bar{q}_f}(\mathbf{p_2})\times \\
    &\nonumber\quad\quad\quad v_{q_f \bar{q}_f} \sigma^{c\bar{c}} _{q_f \bar{q}_f} \delta ^{(4)} (Q-P_1-P_2)\\
    &\nonumber+\int \frac{d^3 p_1}{(2\pi)^3}\frac{d^3 p_2}{(2\pi )^3}f_{g}(\mathbf{p_1}) f_{g}(\mathbf{p_2})\times \\
    &\quad\quad\quad v_{gg} \sigma^{c\bar{c}} _{gg} \delta ^{(4)} (Q-P_1-P_2) ,
\end{align}
where $Q$ is the 4-momentum of the pair, $\sigma^{c\bar{c}}_{q_f \bar{q}_f}$ and $\sigma^{c\bar{c}}_{gg}$ are the matrix elements and $v_{q_f \bar{q}_f}=v_{gg}=\frac{Q^2}{2p_1p_2}$ the relative velocity between incoming particles. The two matrix elements are
\begin{align}
    &\sigma^{c\bar{c}} _{q_f \bar{q}_f} = \frac{2\pi (N_c^2 - 1)\alpha_s^2}{Q^2}v(1-\frac{v}{3})\\
    &\nonumber\sigma^{c\bar{c}} _{gg} = \frac{2\pi (N_c^2 - 1)\alpha_s^2}{Q^2}\times\\
    &~~~~\left[\left( \frac{11}{4} - \frac{3}{2} v^2 + \frac{v^4}{12}\right)\text{artanh}(v) -\frac{v}{24}(59-31v^2) \right],
\end{align}
where $v=\sqrt{1-4\frac{m_c^2}{Q^2}}$. 

It remains to compute the phase-space distributions of quarks and gluons. To do this, we follow previous works on pre-equilibrium photon \cite{Garcia-Montero:2023lrd} and dilepton \cite{Plaschke2025} production in using the effective kinetic theory approach to model the evolution of the distribution functions. These distribution functions facilitate the computation of the energy-momentum tensor encoding macroscopic properties of the medium
\begin{align}
    T^{\mu\nu} = \int \frac{d^3 p}{(2\pi)^3} \frac{p^\mu p^\nu}{p} \left[ \nu_g f_g(\vec{p}) + \nu_q \sum_f \left( f_{q_f}(\vec{p}) + f_{\bar{q}_f}(\vec{p}) \right) \right],
\end{align}
with $\nu_g$ and $\nu_q$ the number of degrees of freedom for gluons and quarks. 

The pre-equilibrium evolution is typically described by conformal Bjorken flow. As discussed in \cite{Giacalone:2019ldn, Du:2020zqg, Du:2020dvp}, it has been demonstrated that many different pre-equilibrium models share certain macroscopic features even well before reaching local thermal equilibrium. This common behavior can be expressed in terms of a single macroscopic dimensionless variable
\begin{align}
    \tilde{w}= \frac{\tau T_{\text{eff}}}{4\pi \eta / s},
\end{align}
with $\tau= \sqrt{t^2-z^2}$ the Bjorken time and the effective temperature $T_{\text{eff}}$ defined via the equilibrium relation between the energy density, which can be computed using the energy-momentum tensor, and the temperature:
\begin{align}
    e(\tau )=: \frac{\pi^2}{30}\nu_{\rm eff} T_{\text{eff}}^4,
    \label{eq:T_eff}
\end{align}
where in QCD kinetic theory for $N_f$ massless flavors $\nu_{\rm eff}=\nu_g +2N_f \frac{7}{8} \nu_q$. The scaling variable $\tilde{w}$ is defined such that the system equilibrates at $\tilde{w}\sim 1$. The universal behavior is visualized in \cite{Giacalone:2019ldn, Du:2020zqg}, which show that, for instance, the local energy density can be expressed only in terms of a dimensionful scale $\left( \tau T^3\right)_{\rm eq}$ and a universal scaling function $\mathcal{E}(\tilde{w})$:  
\begin{align}
    e(\tau)= \frac{\pi^2}{30}\nu_{\text{eff}}\left( \tau T^3\right)_{\rm eq}^{\frac{4}{3}}\frac{\mathcal{E}(\tilde{w})}{\tau ^{\frac{4}{3}}},
    \label{eq:energy_Scaling}
\end{align}
where $\left( \tau T^3\right)_{\rm eq}$, denotes the value of $T(\tau)\tau^3$ of the equilibrated QGP at the end of the pre-equilibrium stage, which, as detailed below, is proportional to the entropy density $(\tau s)_{\rm eq}$ of the thermalized QGP.

While, in practice, QCD kinetic theory simulations are performed for a specific \emph{initial energy density} and coupling strength $\lambda$, the scaling in \eqref{eq:energy_Scaling} implies that the evolution can be inferred for any specific \emph{final state energy density} of interest once the scaling function $\mathcal{E}(\tilde{w})$ is known. This means that, by specifying the physical value of $\left( \tau T^3\right)_{\rm eq}$ at the end of the pre-equilbrium regime, the preceding pre-equilibrium evolution is then completely determined by \eqref{eq:energy_Scaling}. 
Since the shear viscosity of the QGP is small, the late-stage hydrodynamic expansion is approximately isentropic, and the thermal entropy at the onset of hydrodynamic behavior is approximately conserved throughout the hydrodynamic stage and the conversion into particles at freeze-out. 
This allows us to relate $\left( \tau T^3\right)_{\rm eq}$ directly to the measured charged‑particle multiplicity. Indeed, this scale can be computed from \cite{Giacalone:2019ldn}
\begin{align}
    \left( \tau T^3\right)_{\rm eq} = \frac{90}{4\pi^2 \nu_{\text{eff}}} (\tau s(\tau))_{\rm eq}
\end{align}
 %where $(\tau s(\tau))_{\rm eq}$ is proportional to the entropy per unit rapidity, which is a constant of motion for the boost invariant expansion of an ideal fluid. %This assumption becomes increasingly justified as the system evolves towards the regime of hydrodynamics before significant transverse expansion has occurred. 
and the entropy density at thermalization time can be directly estimated from measurements of the charge particle multiplicity $\frac{dN_{ch}}{d\eta}$, as 
% $\frac{dS}{d\eta}$ the entropy per unit of spatial rapidity and $N_{ch}$ the charged particle multiplicity
%  he following equation relates the constant of the motion to quantities specific to the system under investigation:
\begin{align}
    (\tau s(\tau ))_{\text{eq}} =\frac{1}{A_T} \frac{dS}{d\eta}=\frac{1}{A_T}\frac{S}{N_{ch}}\frac{dN_{ch}}{d\eta},
\end{align}
where $A_T$ is the transverse size and $\frac{S}{N_{ch}}=6.7$ \cite{Hanus:2019fnc} denotes the entropy per hadron in a hadron resonance gas at freeze-out. We follow \cite{Coquet:2021lca} and employ $A_T = 104 ~\text{fm}^2$ for the transverse system size of a central Pb+Pb collision, and model the rapidity dependence by varying the charged particle multiplicity $\frac{dN_{ch}}{d\eta}$ according to \cite{ALICE:2016fbt}. We use $\nu_{\text{eff}}=40$ for the high-temperature equation of state.

Beyond the pre-equilibrium evolution of the energy density, the QCD kinetic theory simulations also provide an estimate for the onset of hydrodynamic behavior \cite{Giacalone:2019ldn, Du:2020zqg}, which occurs on time scales 
$$\tilde{w}_{\rm eq}=\tau_{\text{eq}}T_{\text{eq}}/(4\pi \eta/s)\simeq1$$ 
which allows to determine the equilibration time $\tau_{\rm eq}$ and temperature $T_{\rm eq}$ as a function of the shear-viscosity to entropy density $\eta/s$ and the overall energy scale $\sqrt{(\tau T^3)_{\rm eq}}$:

\begin{align}
\label{eq:TauEqTeq}
    &\tau _{\text{eq}} \approx (4\pi \eta/s)^{\frac{3}{2}} \Big/ \sqrt{(\tau T^3)_{\rm eq}} \\
    &T _{\text{eq}} \approx \sqrt{(\tau T^3)_{\rm eq}}(4\pi \eta/s)^{-\frac{1}{2}}
\end{align}

Under the assumption that scaling also works for hard probes in the pre-equilibrium phase, which are produced at earlier times, as it did with photons and dileptons, the charm production rate at one point in phase space, being a dimensionless quantity, should only depend on dimensionless ratios:
\begin{align}
\frac{dN^{c\bar{c}}}{d^4x d^4Q} = \frac{dN^{c\bar{c}}}{d^4x d^4Q}\left(\frac{\tau}{\tau_{eq}},\frac{m_c}{T_{eq}},\frac{Q}{T_{eq}}\right)
\end{align}
Assuming this, any dimensionful integral of $\frac{dN^{c\bar{c}}}{d^4x d^4Q}$ can be written as a scale multiplied by an integral over dimensionless quantities \cite{Plaschke2025,Garcia-Montero:2023lrd}, especially  
\begin{align}
\label{eq:CharmScaling}
    \frac{dN^{c\bar{c}}}{dyd^2x_T}(m_c) = \tau_{\text{eq}}^2 T_{\text{eq}}^4\mathcal{N}\left( \frac{m_c}{T_{\text{eq}}}\right),
\end{align}
with the charm-scaling function
\begin{align}
    \nonumber\mathcal{N}\left( \frac{m_c}{T_{\text{eq}}}\right) = \int d \frac{\tau}{\tau_{\text{eq}}}\frac{\tau}{\tau_{\text{eq}}}\int d\zeta \int d^2 \frac{q_T}{T_{\text{eq}}} \int d\frac{M}{T_{\text{eq}}} \frac{M}{T_{\text{eq}}}\\
    \frac{dN^{c\bar{c}}}{d^4x d^4Q}\left(\frac{\tau}{\tau_{eq}},\frac{m_c}{T_{eq}},\frac{Q}{T_{eq}}\right),
\end{align}
where $\zeta$ is the space-time rapidity and $M$ is the invariant mass associated with $Q$. Such scaling is expected to work if the production is dominated by late times when the system has lost its memory of the initial conditions, such as is the case for photons and dileptons. But since charm quarks are  not only heavy, but can also be produced from gluon processes at leading order, the heavy-flavor production is expected to extend to earlier times, where the initial conditions might still have large effects. We will test this universality, by performing QCD kinetic theory simulations at different couplings $\lambda = 5, 10, 20$, which upon performing the scale setting as indicated above, mainly corresponds to a variation of the initial conditions and early time dynamics. 

Besides considering QCD kinetic theory simulations at different coupling strength, we will also compare our results to those from a simple model, where the evolution of the phase-space distribution is parametrized in terms of modified equilibrium quark and gluon distributions to describe the pre-equilibrium features of quark suppression and momentum anisotropy \cite{Coquet:2021lca}:
\begin{align}
    &f_g (\tau, p_T, p_z)=f_{BE}\left( \frac{\sqrt{p_T^2+\xi (\tau) p_z^2}}{\Lambda (\tau )}\right)\\
    &f_{q/\bar{q}} (\tau, p_T, p_z)=q(\tau )f_{FD}\left( \frac{\sqrt{p_T^2+\xi (\tau) p_z^2}}{\Lambda (\tau )}\right)
\end{align}
Since the phase-space distributions are of the Romatschke-Strickland form, we will refer to this as Romatschke-Strickland (RS) model, where $\xi (\tau ) \ge 1$ is the momentum anisotropy, $q(\tau )\in [0,1]$ the quark suppression factor and $\Lambda(\tau )$ the effective transverse temperature. The functions $\xi (\tau )$, $q(\tau )$ and $\Lambda(\tau )$ are determined by matching the evolution of the energy-momentum tensor of this model to the QCD kinetic theory evolution (see \cite{Coquet:2021lca} for details). Since this requirement ensures exact evolution according to the same hydrodynamic attractor scaling, regardless of the parameter choice ($\eta/s$ in $\tilde{w}$ is the only parameter), the scaling of the charm-production in \eqref{eq:CharmScaling} is guaranteed to be exact in this model. 

%The details of this identification between the Romatschke-Strickland and the EKT attractors are thoroughly explained in \cite{Coquet:2021lca}.

\section{Instantaneous Production Rate}

\begin{figure*}[t]
    \centering
    \includegraphics[width=0.48\textwidth]{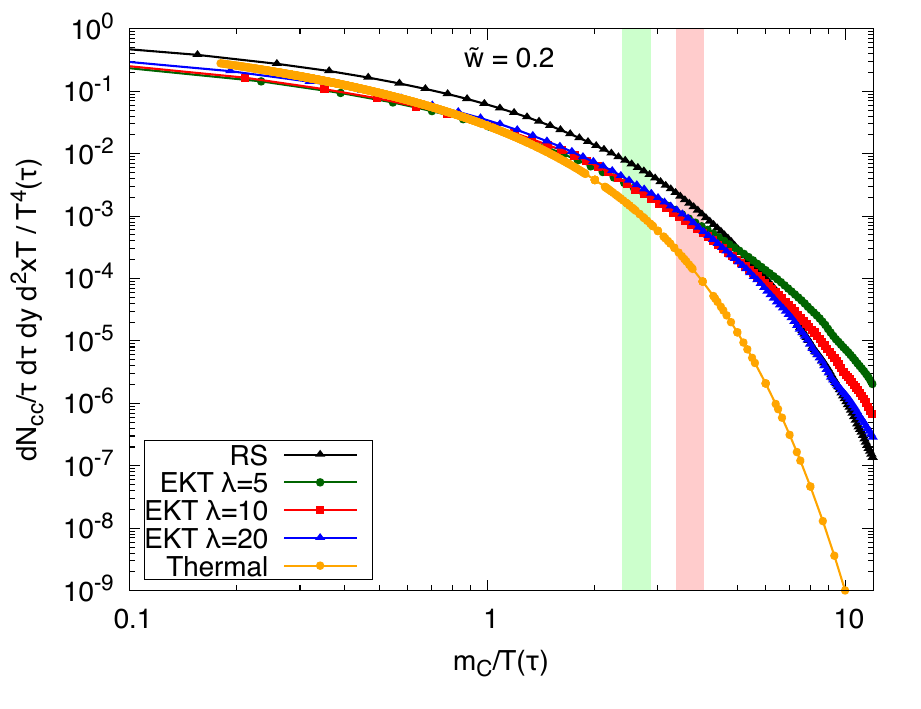}
    \includegraphics[width=0.48\textwidth]{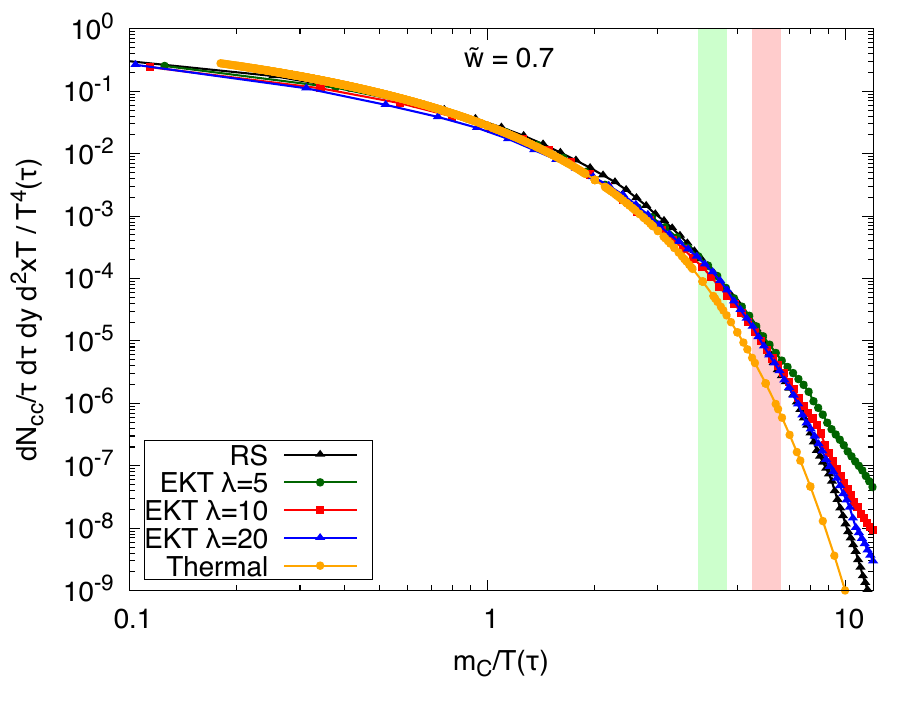}
    \includegraphics[width=0.48\textwidth]{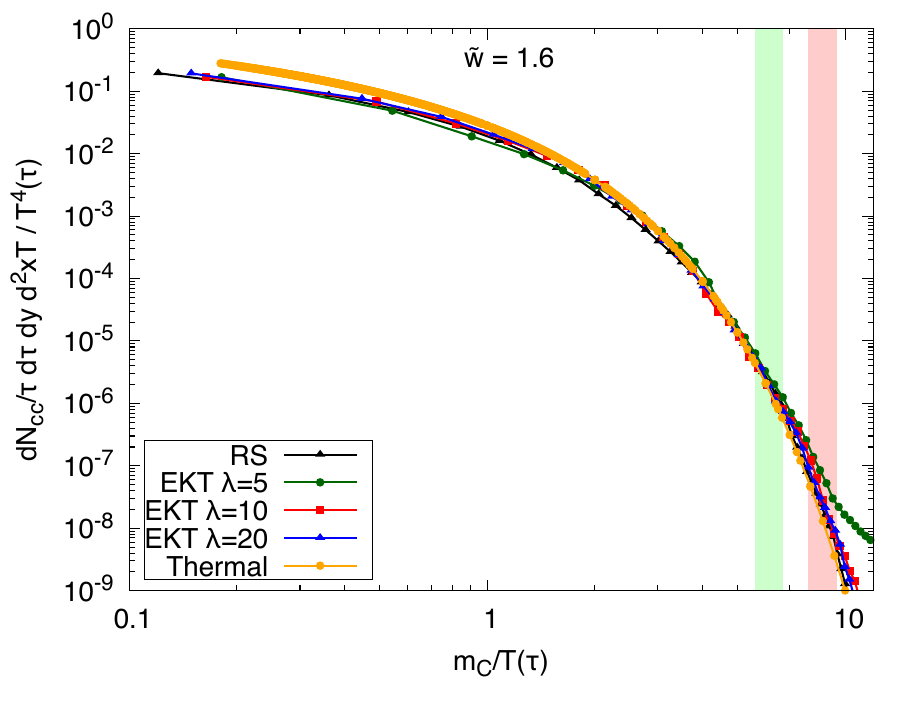}
    \includegraphics[width=0.48\textwidth]{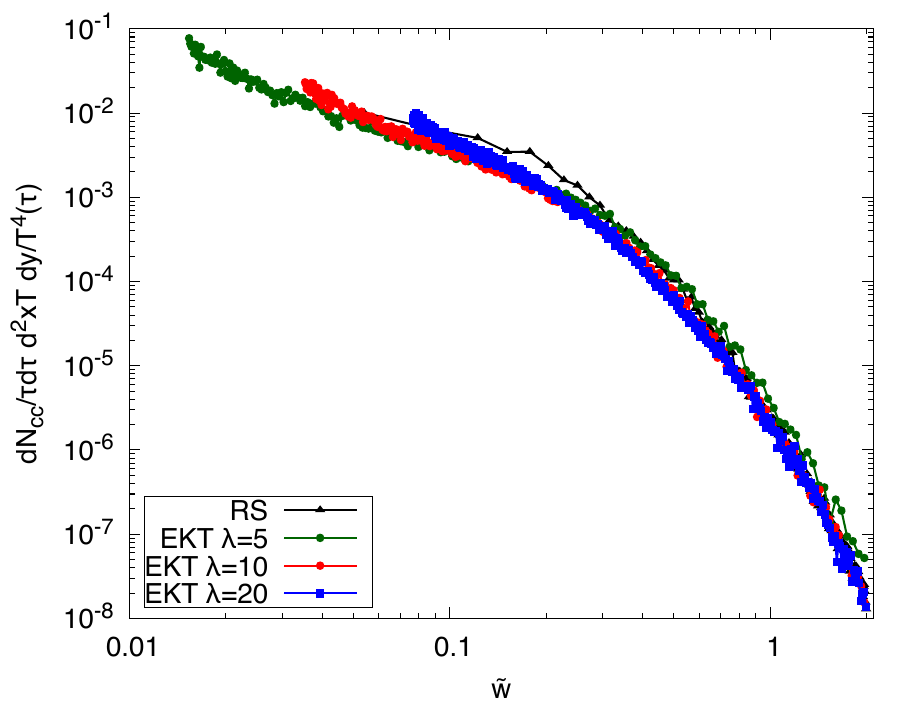}
    \caption{Scaling functions for the instantaneous charm production rate, (top and bottom left) as a function of $\frac{m_c}{T(\tau )}$, and (bottom right) as a function of $\Tilde{w}$. Calculations from QCD kinetic theory (EKT) are shown for different values of coupling $\lambda$, and compared with results from the RS model, and thermal distributions. Green and red boxes show physical ranges for $
\frac{m_c}{T(\tau )}$ with $1.25~{\rm GeV} < m_c<1.5~{\rm GeV}$, for $\eta/s=0.16$ and $\eta/s=0.32$, respectively.}
    \label{fig:InstScaling}
\end{figure*}

In order to characterize the pre-equilibrium production of heavy $q\bar{q}$ pairs, we first consider the instantaneous production rate $\frac{dN^{c\bar{c}}}{\tau d\tau dy d^2x_T}$. Based on the above arguments, the scaling function for this rate can be extracted as
\begin{align}
    \frac{dN^{c\bar{c}}}{\tau d\tau dy d^2x_T}=T^4_{\text{eff}}(\tau)\mathcal{I}\left( \frac{m_c}{T_{\text{eff}}(\tau )}\right)
\end{align}
We present our results for the instantaneous production rate in Fig. \ref{fig:InstScaling}, where the different panels show the scaling functions $\frac{dN^{c\bar{c}}}{\tau d\tau dy d^2x_T}/T^4_{\text{eff}}(\tau)$ for three different times $\tilde{w}=0.2,~ 0.7, ~ 1.6$. Different curves show the results for QCD kinetic theory at different coupling strength and the Romatschke-Strickland model. In addition, these results are compared to the equilibrium rate, where the gluon- and quark-distribution functions are Bose-Einstein and Fermi-Dirac distributions. We indicate the physical ranges for $
\frac{m_c}{T(\tau )}$ as colored vertical boxes, with the green box covering $1.25~{\rm GeV} < m_c<1.5~{\rm GeV}$ for $\eta/s=0.16$ and the red box covering the same mass range for $\eta/s=0.32$, which as discussed in the previous section, affects the determination of $T_{\rm eff}(\tilde{w})$.\footnote{The physical effective temperature at $\tilde{w}$ can be computed using the scaling of the energy density \eqref{eq:energy_Scaling}.} One observes that at early times $\tilde{w}=0.2$, the two non-equilibrium models differ substantially, as the production of heavy-quark pairs is still sensitive to the initial conditions. As expected, these differences become less pronounced at large values of $\tilde{w}$, when the phase-space distributions of QCD kinetic theory approach the Romatschke-Strickland model. Similarly, all models converge towards the thermal charm production rate at $\tilde{w}=1.6$ where hydrodynamics is applicable to describe the sub-sequent evolution. However, it is important to realize that as  time $\tilde{w}$ increases, the effective temperature falls, causing the physical values of $\frac{m_c}{T_{\text{eff}}}$ to increase. We also note that for large values of $\frac{m_c}{T_{\text{eff}}}>10$, the different models do not show a universal behavior even at late times,  suggesting that universal scaling function are ill-suited to describe bottom quark production, which if relevant at all, is always sensitive to the initial conditions. Next, in order to compare the charm production at different times, the bottom-right panel in Fig.\ref{fig:InstScaling} depicts the scaled instantaneous rate against $\tilde{w}$, for a charm quark mass of $1.25$ GeV at $\frac{\eta}{s}=0.32$. Early time deviations between the QCD kinetic results at different coupling strength $\lambda$ reflect the dependence on the initial conditions of the evolution. However, all curves demonstrate quick convergence to a scaling function, which features sizeable production rates at early times $\tilde{w} \lesssim 0.3$, followed by a rapid decay at later times.

\section{Integrated Charm Yield}
\begin{figure}[!h]
    \centering
    \includegraphics[width=1.0\linewidth]{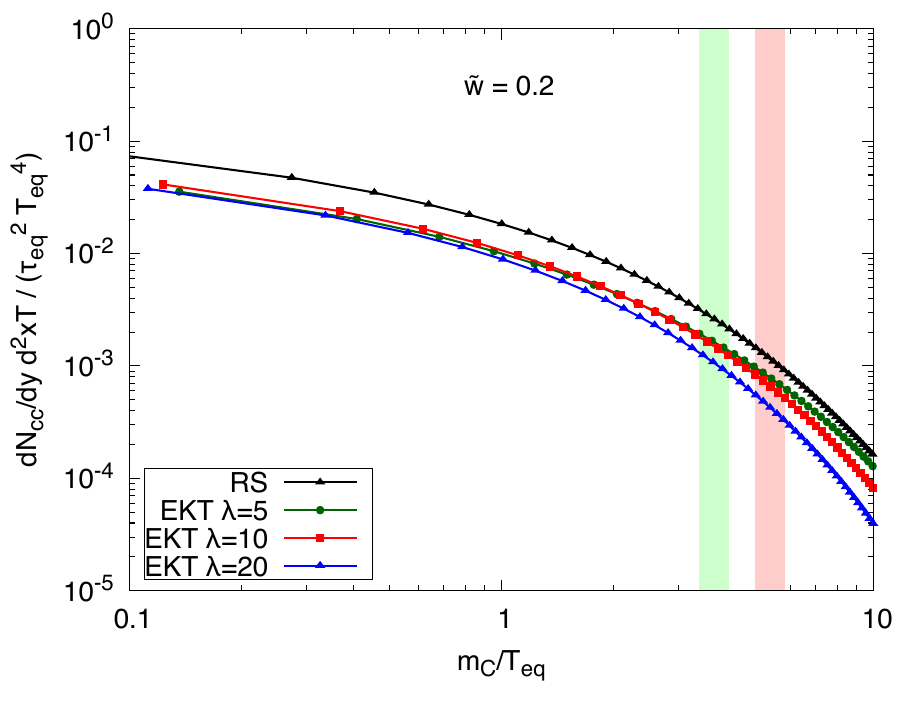}
    \includegraphics[width=1.0\linewidth]{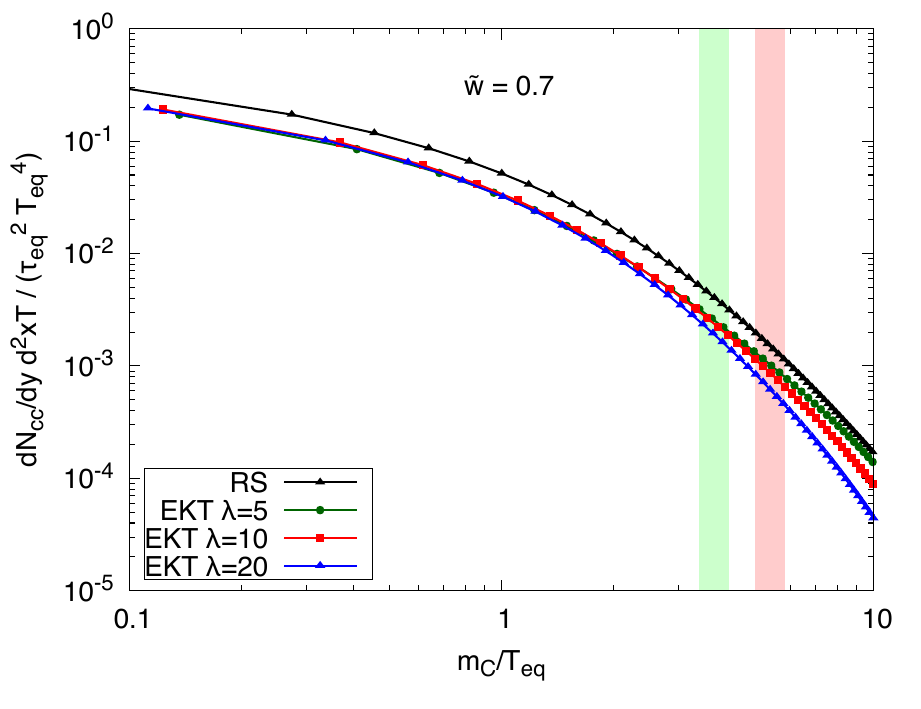}
    \includegraphics[width=1.0\linewidth]{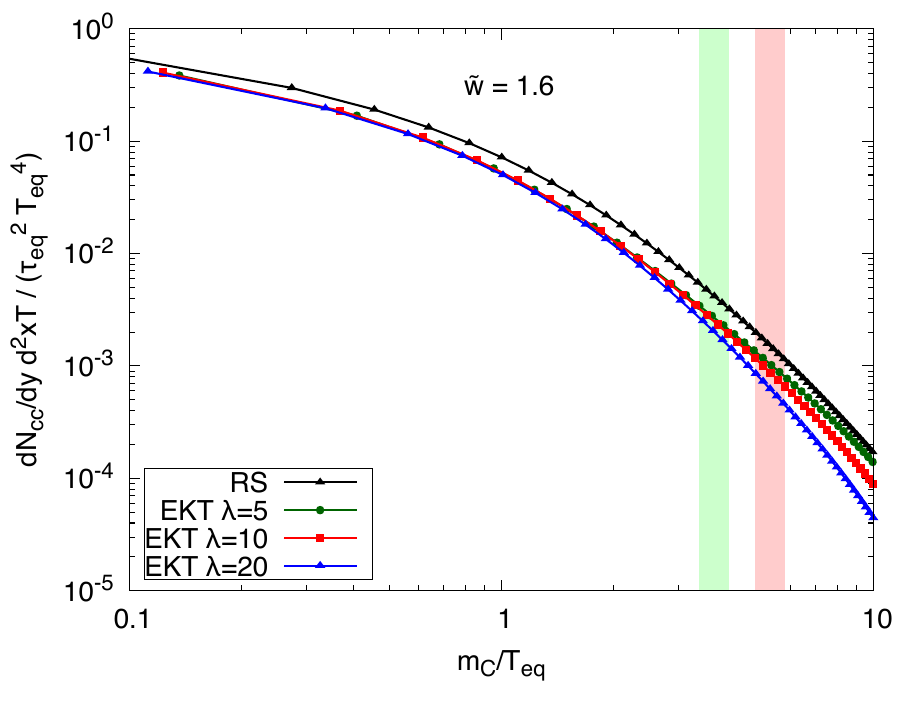}
    \caption{Scaling functions for the integrated charm yield produced during the pre-equilibrium evolution as a function of $\frac{m_c}{T_{\rm eq}}$. Calculations from QCD kinetic theory (EKT) are shown for different values of $\lambda$, and compared with RS model. Green and red boxes show ranges for $
\frac{m_c}{T_{\rm eq}}$ with $1.25~{\rm GeV} < m_c<1.5~{\rm GeV}$, for $\eta/s=0.16$ and $\eta/s=0.32$, respectively.}
    \label{fig:YieldScaling}
\end{figure}

By integrating the instantaneous production rate $\frac{dN^{c\bar{c}}}{\tau d\tau dy d^2x_T}$ up to a certain time, we obtain the differential yield $\frac{dN^{c\bar{c}}}{dyd^2x_T}(m_c)$ of $c\bar{c}$ pairs per unit rapidity and transverse area, produced up to this point, which is expected to satisfy the scaling in 
\eqref{eq:CharmScaling}. We present our results for this integrated charm yield, in
Fig.~\ref{fig:YieldScaling}, where the different panels contains the scaling functions
for three different upper time limits $\tilde{w}=0.2,~0.7,~1.6$. Colored boxes in Fig.~\ref{fig:YieldScaling} indicate physical values of $m_c/T_{\rm eq}$ for a charm quark mass $1.25~{\rm GeV} <m_c< 1.5~{\rm GeV}$, with green boxes corresponding to $\eta/s=0.16$ and red boxes to $\eta/s=0.32$, which according to \eqref{eq:TauEqTeq} result in different temperatures $T_{\rm eq}$ at the time when the QGP equilibrates. One observes that the scaling for the integrated charm yield improves at later times, especially for low $\frac{m_c}{T_{\text{eq}}}$, but there is still significant disagreement between the results for masses around and above the charm mass. This is due to the fact that charm production occurs pre-dominantly at early times, where the initial conditions still play a significant role. The production yield predicted by the Romatschke-Strickland model is larger. For low masses, this is consistent with the low $\tilde{w}$ limit of the instantaneous rate, where the bulk of the production occurs. Another effect that causes this pattern is that, unlike in QCD kinetic theory, where the simulation starts at an initial time $\tau = \frac{1}{Q_s}$, the Romatschke-Strickland simulation effectively starts at $\tau = 0$. This explains why the Romatschke-Strickland result is larger even at high masses and implies that it is likely to be an overestimate of the pre-equilibrium charm yield. Furthermore, this initial time protects the calculation from double counting of pre-equilibrium charm production and charm production from initial hard scatterings associated with the typical time scale of $\tau \approx \frac{1}{m_c}$.

\begin{figure}[t]
    \centering
    \includegraphics[width=1.0\linewidth]{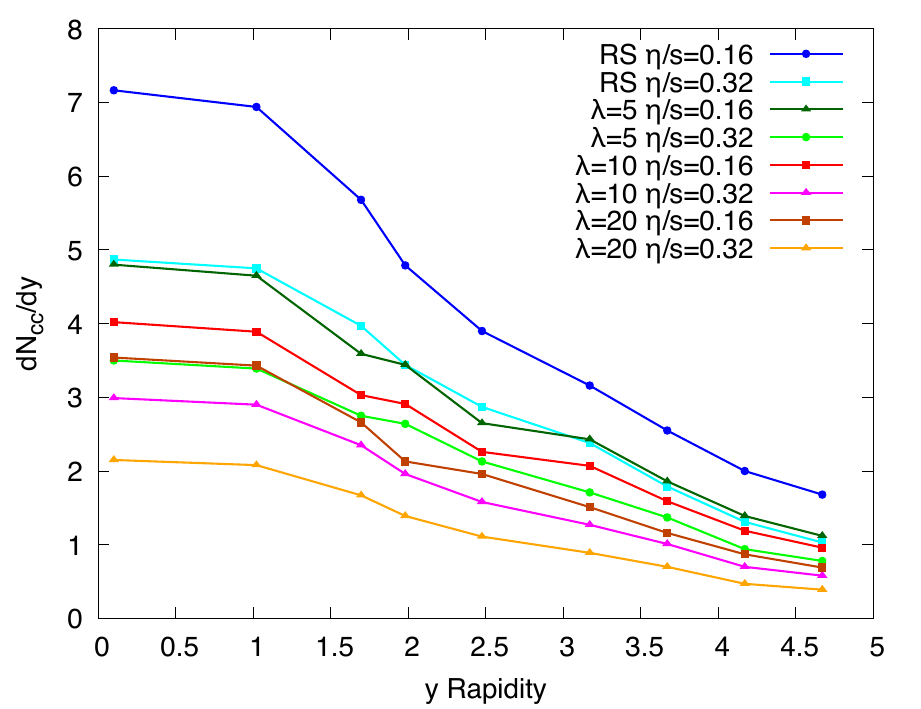}
    \caption{Rapidity distribution of the total charm production per unit rapidity $dN^{c\bar{c}}/dy$ during the pre-equilibrium phase for $\eta/s=0.16$ and $\eta/s=0.32$, computed with QCD kinetic theory varying the coupling $\lambda$, and with RS model.}
    \label{fig:Rapidity-Distribution}
\end{figure}

Based on the scaling functions for the integrated charm production in Fig.~\ref{fig:YieldScaling}, it is then straightforward to obtain the differential yield $dN^{c\bar{c}}/dy$ by integrating over transverse area. 
Fig. \ref{fig:Rapidity-Distribution} displays the rapidity distribution of the pre-equilibrium yield predicted by each model for two distinct physical setups with $m_c=1.5$ GeV. Since all models used in this work assume boost invariance, the rapidity dependence is obtained by mapping the charged particle multiplicity onto the rapidity~\footnote{This simplification is not expected to be critical, since the observables are sensitive to momentum space and not coordinate space distributions. }. This correspondence is established using data for charged particle multiplicities at different rapidities for collisions in the 0-5\% centrality class from ALICE \cite{ALICE:2016fbt}. More specifically, for any value of the rapidity $y$, the corresponding charged particle multiplicity can be used to compute $(\tau T^3)_{\rm eq}$. This in combination with a value for $\frac{\eta}{s}$ then leads to values for $T_{\text{eq}}$, $\tau_{\text{eq}}$ and finally $\frac{m_c}{T_{\text{eq}}}$, which can be used to obtain the predicted charm yield for this physical scenario from the scaling functions. The results confirm the expected shape of the rapidity distribution. As in the case of dileptons \cite{Coquet:2021lca}, lower $\frac{\eta}{s}$ leads to larger production. The predicted pre-equilibrium charm yield must now be compared with the charm yield coming from initial hard scatterings of the collision.

\section{Charm production from initial hard scatterings}
The inclusive charm production cross section from initial hard scatterings in perturbative QCD based on collinear factorisation exhibits large theoretical uncertainties~\cite{Cacciari:1998it,Kniehl:2012ti}. In particular, the uncertainties evaluated with the variations of the factorisation and renormalisation scales and associated to missing higher order corrections are  dominating for the transverse momentum integrated cross sections and reach a variation of around 5 between the lower bound and the higher bound of the uncertainty. A reduction of the scale variation uncertainties comparing NLO with NNLO calculation is observed, but the uncertainties remain by far larger than the experimental uncertainties~\cite{Catani:2020kkl,dEnterria:2026tuz}.  In collisions involving nuclei as well as in extreme kinematics, large uncertainties on the gluon parton distribution function are relevant contributions to the uncertainty budget as well~\cite{Ethier:2020way,Apolinario:2022vzg,Klasen:2023uqj}.

\begin{figure}
    \centering
    \includegraphics[width=1.0\linewidth]{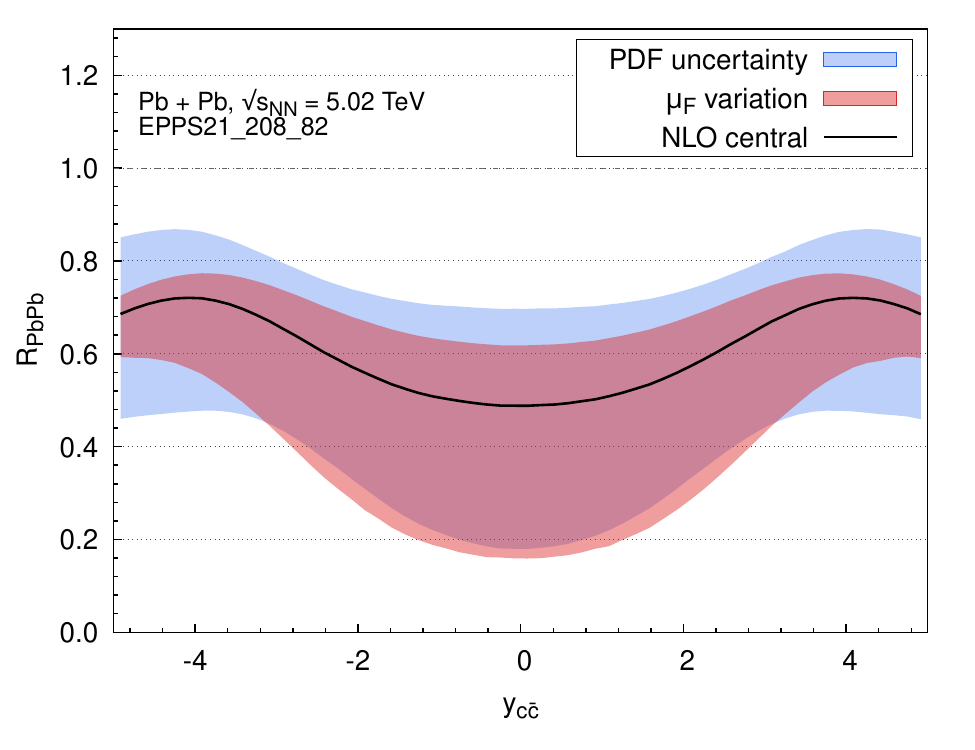}
    \caption{Nuclear modification factor for $c\Bar{c}$ production computed at next-to-leading order with the MadGraph software~\cite{Alwall:2014hca}, as a function of the rapidity of the $c\Bar{c}$ pair.}
    \label{fig:RAA}
\end{figure}

%Based on these considerations, the following estimates are  based on experimental data at the LHC. 
The production yield in PbPb collisions is estimated based on the nuclear modification factor $R_{AA} = \frac{d\sigma_{c\bar{c}}^{PbPb}/dy}{A^2 d\sigma_{c\bar{c}}^{pp}/dy}$, where $\langle T_{AA}\rangle$ is the nuclear overlap of a given centrality class. The $R_{AA}$ is evaluated with the MadGraph software~\cite{Alwall:2014hca} at fixed-order NLO, with a charm quark mass of $m_c = 1.55 ~\mathrm{GeV}/c^2$, and with the nuclear parton distribution functions from EPPS21~\cite{Eskola:2021nhw}. %This procedure neglects the centrality dependence of the nuclear modification of parton distribution functions. %\textcolor{red}{I put only EPPS21 for now ...}
%\cite{AbdulKhalek:2022fyi,Eskola:2016oht} \textcolor{red}{didn't find ref to NCTEQ newest release, I am lost in the large number of fits from their group...Maurice do you know the best ref?}. 
These nuclear parton distribution functions assume that inclusive hard particle production in proton-nucleus collisions can be used to deduce the nuclear modification of parton distribution functions, see for a detailed discussion in~\cite{Apolinario:2022vzg}. The resulting calculation provides a fully inclusive production cross section in lead-lead collisions. The resulting nuclear modification factor is shown as a function of the rapidity of the $c\Bar{c}$ pair in Fig.\ref{fig:RAA}. The displayed uncertainties are the PDF uncertainties as well as an uncertainty from the variation of the factorization scale $\mu_F$ by factors $[0.5; 1; 2]$. The uncertainty due to the variation of the renormalization scale cancel in the calculation.

Next, to get an estimate of the yield in PbPb collisions, the nuclear modification factor must be multiplied by the $c\Bar{c}$ production cross section in proton-proton collision. It is evaluated by using the cross section in proton-proton collisions at midrapidity $d\sigma_{c\bar{c}}^{pp}/dy|_{y\in[-0.5,0.5]}$ at $\sqrt{s_{NN}}=5.02$~TeV measured by the ALICE collaboration, including the experimental measurement of several charm baryon species~\cite{ALICE:2023sgl}, to set the absolute production scale at midrapidity. The rapidity distribution of charm production in proton-proton collisions is taken from MadGraph simulations, using the proton PDF set CT18ANLO~\cite{Hou:2019efy}, which is the reference used by EPPS21. The uncertainties on scale variations are not evaluated since the experimental data is used as anchor point. Hence, it is assumed for simplicity that the scale uncertainties are fully correlated as function of rapidity.

In order to estimate the production in ranges of centrality, we do not take into account the centrality dependence of nuclear modifications of parton distribution functions and assume scaling with the nuclear overlap function~\cite{Miller:2007ri}. This simplification yields to a overestimation of the expected charm yield from initial hard scatterings in central collisions since the charm production is suppressed due to the suppression of gluons in the parton distribution functions and since the nuclear effects are expected to be emphasized in central collisions. The $c\Bar{c}$ production yield in PbPb collisions is thus estimated as $dN_{c\bar{c}}^{PbPb}/dy = R_{AA} \cdot \langle T_{AA} \rangle \cdot d\sigma_{c\bar{c}}^{pp}/dy$, where the nuclear overlap $\langle T_{AA}\rangle$ for 0-5\% central PbPb collisions is taken from Monte Carlo Glauber predictions~\cite{Loizides:2017ack}.

%In Fig.~\ref{fig:Yield_initialscattering}, we show the charm production from initial scattering in the 0-5\% centrality bin. We calculate a production yield of charm in the rapidity bin $y\in[-0.5,0.5]$ of. \textcolor{red}{compare with measurement}. 

%MW: to do, explain that we take absolute scale from pp mesaurement of ALCE. rapidity shape from theory. 

%Need to take something for RHIC? 

%RAA from Madgraph for most cenrtal collisiosn (0-10 percent )

%General introduction of process diagrams.

%Mention that total charm well measured by ALICE in pp at 5 TeV and potential to do the same in LHCb.

%Mention main uncertainties including nuclear PDF and factorisation scale uncertainties and centrality dependence.

%Say that we don't consider the centrality dependence for the moment. 

%Show plot with uncertainties for RAA with 3 different pDFs. 

\section{Total Charm production}

\begin{figure}
    \centering
   \includegraphics[width=1.0\linewidth]{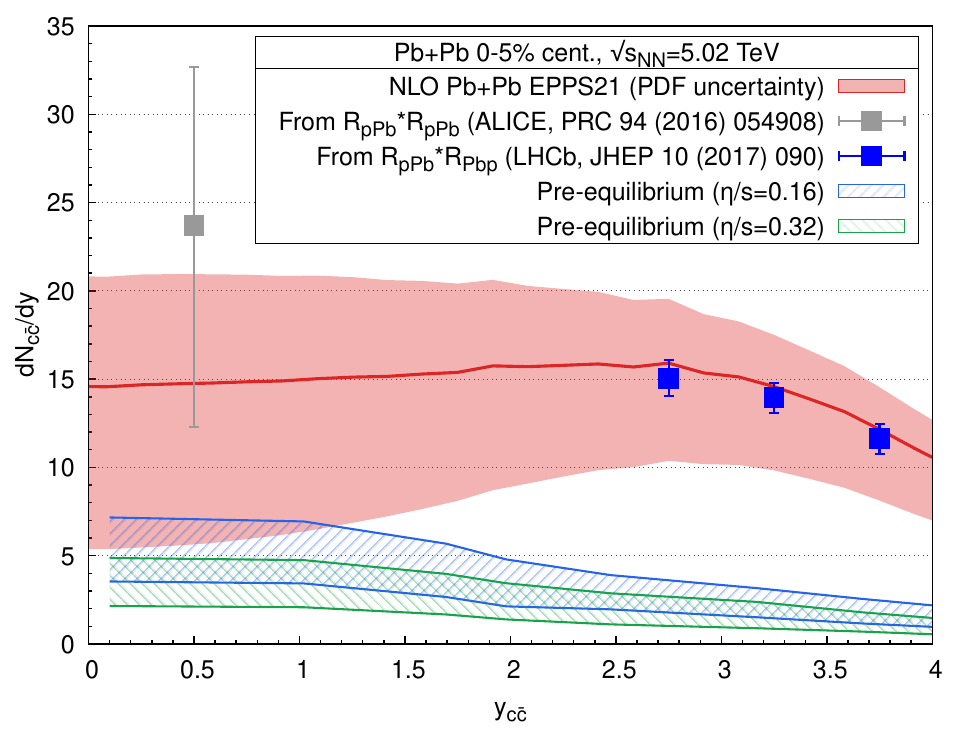}
    \caption{Comparison of charm production yields from initial hard scattering, and from pre-equilibrium for two values of $\eta/s$. The uncertainty band associated to the pre-equilibrium production correspond to different values of $\lambda$ in the EKT calculation, or the usage of RS model. The uncertainty band on the initial hard scattering calculation represents the PDF uncertainty of EPPS21.}
    \label{fig:Yield_comparison}
\end{figure}

Figure~\ref{fig:Yield_comparison} presents a comparison between the $c\bar{c}$ production yield obtained from initial hard scatterings, as computed in the previous section, and the one arising from the pre-equilibrium stage. The shaded bands associated with the pre-equilibrium contribution represent the envelope of the curves shown in Fig.~\ref{fig:Rapidity-Distribution}, corresponding separately to the two values of $\eta/s = 0.16$ and $0.32$.

The uncertainty band on the initial hard-scattering yield reflects the PDF-induced uncertainty in the nuclear modification factor shown in Fig.~\ref{fig:RAA}, which dominates over the scale-variation uncertainty across most of the rapidity interval.

Due to the substantial magnitude of this uncertainty, the pre-equilibrium contribution at $y_{c\bar{c}} = 0$ may account for a fraction of the total charm yield ranging from approximately 10\% up to a value comparable to the yield from initial hard scatterings. Comparing the pre-equilibrium contribution to the central value of the initial hard-scattering estimate indicates a contribution at the level of approximately 10–50\%. In this context, we expect positive corrections from the missing higher-order contributions to our leading-order charm-quark production calculation, as observed in vacuum~\cite{Nason:1989zy,dEnterria:2026tuz}. Therefore, the calculation presented here is expected to provide a lower bound for higher-order calculations based on the same setup.
Beyond higher order perturbative corrections, there are also non-perturbative effects, in particular the so called Sommerfeld correction associated with the attractive/repulsive interaction of color singlet/octet $Q\bar{Q}$ pairs during over the course of the production process~\cite{Bodeker:2012zm}. While a precision calculation should take such effects into account, equilibrium calculations in~\cite{Bodeker:2012zm} indicate that the net effect is small due to cancellations between enhancement and suppression in the color singlet and octet channels.

Given the large theoretical uncertainties on $R_{AA}$ for production from initial hard scatterings evaluated based on the theoretical uncertainties, we provide alternatively a data-driven approach.  Instead of using the theoretical calculation for the estimate of the charm production yield, we can also assume that any effect visible in the proton-nucleus data is also transferred to the nucleus-nucleus collisions and that the $R_{AA}$ in absence of any additional effects can be calculated based on the $R_{pA}$ at the same centre of mass energy as:
\begin{align}
    R_{AA}^{initial} (y=y_0) \approx R_{pA}(y=y_0) \cdot R_{pA} (y= -y_0) 
\end{align}
This approach was first introduced in~\cite{ALICE:2015sru}. For the estimate of the $R_{AA}^{initial}$, we take for simplicity the measurements of $D^0$ assuming that the uncertainty of the hadronization fraction modifications between proton-proton and proton-lead are not introducing a modification of the rapidity distribution of the estimated nuclear modification. We then estimate the yield in lead-lead collisions analogously as before, by multiplying $R_{AA}^{initial}$ with the same proton-proton cross section also used for the computation of the pQCD estimate. The corresponding points are shown in gray for ALICE~\cite{ALICE:2016yta} and in blue for LHCb~\cite{LHCb:2017yua}, and the error bars are computing by propagating the total uncertainty of the $R_{pA}$ measurement. The points from LHCb indicate that a precision measurements in nucleus-nucleus collisions comparable to the one in proton-lead and proton-proton collisions should allow to extract the charm production from the pre-equilibrium as the difference between the yield estimated from proton-proton and proton-nucleus data and the measurement in nucleus-nucleus collisions.

%Explain overall strategy to compare measured charm with pp extrapolation.

%Make an error estimate.

%Make an example for central data point. 

%{\it Parameter estimate from current data. ---} 

%Explain overall strategy to compare measured charm with pp extrapolation.

%Make an error estimate.

%Make an example for central data point. 

\section{Conclusion}
The charm production from the pre-equilibrium phase of heavy-ion collisions at the LHC prior to thermalization is a relevant fraction of the charm production yield. Pre-equilibrium charm production is highly complementary to dilepton production as the dominant production occurs through gluon fusion. Since the pre-equilibrium production is largest at early times, we find that the yield does not obey the same universal scaling as dilepton charm production, instead a remaining sensitivity to the initial conditions persists as we exemplified when considering variations of the effective kinetic theory description for different couplings $\lambda$. The overall size of the production yield turns out to be relevant for phenomenology with a size of 10\% or larger depending on the rapidity. An extraction of the pre-equilibrium charm production is so far not possible due to the large uncertainties of total charm production in nucleus-nucleus collisions. Assuming that nuclear modification in proton-nucleus collisions is sufficiently well understood, precision measurements of the total charm production yield in nucleus-nucleus collisions with similar precision as in proton-proton collisions and proton-nucleus collisions will allow the extraction of the pre-equilibrium charm production. Charm precision production measurements in nucleus-nucleus collisions are a major goal for the LHCb Upgrade 2 and ALICE 3 foreseen for LHC Run 5.   This perspective opens a new access to the thermalization process in heavy-ion collisions.

\section*{Acknowledgements}
%This work is supported in part by the Deutsche Forschungsgemeinschaft (DFG, German Research Foundation) through the CRC-TR 211 ``Strong-interaction matter under extreme conditions'' project number 315477589 – TRR 211 and in part in the framework of the GLUODYNAMICS project funded by the ``P2IO LabEx (ANR-10-LABX-0038)'' in the framework ``Investissements d'Avenir'' (ANR-11-IDEX-0003-01) managed by the Agence Nationale de la Recherche (ANR), France.

This work is supported in part by the Deutsche Forschungsgemeinschaft (DFG, German Research Foundation) through the CRC-TR 211 ``Strong-interaction matter under extreme conditions'' project number 315477589 – TRR 211, and in part in the framework the Agence Nationale de la recherche Grant AccessEmergence (No. ANR-23-CE31-0013). %. % France, and in part by Xunta de Galicia (Centro singular de investigacion de
%Galicia accreditation 2019-2022), European Union ERDF,
%the “Maria de Maeztu” Units of Excellence program under
%project CEX2020-001035-M. %, the Spanish Research State
%Agency under project PID2020-119632GB-I00, and European Research Council under project ERC-2018-ADG835105 YoctoLHC.
This work is supported by a public grant led by the National Research Agency (ANR) as part of the "Investments for the Future" program, through the "ADI 2023" project funded by IDEX Paris-Saclay, ANR-11-IDEX-003".

\bibliographystyle{elsarticle-num} 
\bibliography{biblio}

@article{Apolinario:2022vzg,
    author = "Apolin\'ario, L. and Lee, Y. J. and Winn, M.",
    title = "{Heavy quarks and jets as probes of the QGP}",
    eprint = "2203.16352",
    archivePrefix = "arXiv",
    primaryClass = "hep-ph",
    doi = "10.1016/j.ppnp.2022.103990",
    journal = "Prog. Part. Nucl. Phys.",
    volume = "127",
    pages = "103990",
    year = "2022"
}

@article{Andronic:2025jbp,
     author = "Andronic, Anton and Arnaldi, Roberta",
    title = "{Quarkonia and Deconfined Quark{\textendash}Gluon Matter in Heavy-Ion Collisions}",
    eprint = "2501.08290",
    archivePrefix = "arXiv",
    primaryClass = "nucl-ex",
    doi = "10.1146/annurev-nucl-121423-101041",
    journal = "Ann. Rev. Nucl. Part. Sci.",
    volume = "75",
    number = "1",
    pages = "351--375",
    year = "2025"
}

@article{Andronic:2006ky,
    author = "Andronic, A. and Braun-Munzinger, P. and Redlich, K. and Stachel, J.",
    title = "{Statistical hadronization of heavy quarks in ultra-relativistic nucleus-nucleus collisions}",
    eprint = "nucl-th/0611023",
    archivePrefix = "arXiv",
    primaryClass = "nucl-th",
    doi = "10.1016/j.nuclphysa.2007.02.013",
    journal = "Nucl. Phys. A",
    volume = "789",
    pages = "334--356",
    year = "2007"
}

@article{Rapp:2018qla,
    author = "Rapp, R. and Gossiaux, P. B. and Andronic, A. and Averbeck, R. and Masciocchi, S. and Beraudo, A. and Bratkovskaya, E. and Braun-Munzinger, P. and Cao, S. and Dainese, A. and others",
    title = "{Extraction of Heavy-Flavor Transport Coefficients in QCD Matter}",
    eprint = "1803.03824",
    archivePrefix = "arXiv",
    primaryClass = "nucl-th",
    doi = "10.1016/j.nuclphysa.2018.09.002",
    journal = "Nucl. Phys. A",
    volume = "979",
    pages = "21--86",
    year = "2018"
}

@article{Ollitrault:2007du,
    author = "Ollitrault, J. Y.",
    title = "{Relativistic hydrodynamics for heavy-ion collisions}",
    eprint = "0708.2433",
    archivePrefix = "arXiv",
    primaryClass = "nucl-th",
    doi = "10.1088/0143-0807/29/2/010",
    journal = "Eur. J. Phys.",
    volume = "29",
    pages = "275--302",
    year = "2008"
}

@article{Uphoff:2010sh,
    author = "Uphoff, J. and Fochler, O. and Xu, Z. and Greiner, C.",
    title = "{Heavy quark production at RHIC and LHC within a partonic transport model}",
    eprint = "1003.4200",
    archivePrefix = "arXiv",
    primaryClass = "hep-ph",
    doi = "10.1103/PhysRevC.82.044906",
    journal = "Phys. Rev. C",
    volume = "82",
    pages = "044906",
    year = "2010"
}

@article{Zhang:2007yoa,
    author = "Zhang, B. W. and Ko, C. M. and Liu, W.",
    title = "{Thermal charm production in a quark-gluon plasma in Pb-Pb collisions at s**(1/2)(NN) = 5.5-TeV}",
    eprint = "0709.1684",
    archivePrefix = "arXiv",
    primaryClass = "nucl-th",
    doi = "10.1103/PhysRevC.77.024901",
    journal = "Phys. Rev. C",
    volume = "77",
    pages = "024901",
    year = "2008"
}

@article{Zhou:2016wbo,
    author = "Zhou, K. and Chen, Z. and Greiner, C. and Zhuang, P.",
    title = "{Thermal Charm and Charmonium Production in Quark Gluon Plasma}",
    eprint = "1602.01667",
    archivePrefix = "arXiv",
    primaryClass = "hep-ph",
    doi = "10.1016/j.physletb.2016.05.051",
    journal = "Phys. Lett. B",
    volume = "758",
    pages = "434--439",
    year = "2016"
}

@article{Song:2024hvv,
    author = "Song, T. and Grishmanovskii, I. and Soloveva, O. and Bratkovskaya, E.",
    title = "{Thermal production of charm quarks in relativistic heavy-ion collisions}",
    eprint = "2404.00425",
    archivePrefix = "arXiv",
    primaryClass = "nucl-th",
    doi = "10.1103/PhysRevC.110.034906",
    journal = "Phys. Rev. C",
    volume = "110",
    pages = "034906",
    year = "2024"
}

@article{Busza:2018rrf,
    author = "Busza, W. and Rajagopal, K. and van der Schee, W.",
    title = "{Heavy Ion Collisions: The Big Picture, and the Big Questions}",
    eprint = "1802.04801",
    archivePrefix = "arXiv",
    primaryClass = "hep-ph",
    doi = "10.1146/annurev-nucl-101917-020852",
    journal = "Ann. Rev. Nucl. Part. Sci.",
    volume = "68",
    pages = "339--376",
    year = "2018"
}

@article{Schlichting:2019abc,
    author = "Schlichting, S. and Teaney, D.",
    title = "{The First fm/c of Heavy-Ion Collisions}",
    eprint = "1908.02113",
    archivePrefix = "arXiv",
    primaryClass = "nucl-th",
    doi = "10.1146/annurev-nucl-101918-023825",
    journal = "Ann. Rev. Nucl. Part. Sci.",
    volume = "69",
    pages = "447--476",
    year = "2019"
}

@article{Berges:2020fwq,
    author = "Berges, J. and Heller, M. P. and Mazeliauskas, A. and Venugopalan, R.",
    title = "{QCD thermalization: Ab initio approaches and interdisciplinary connections}",
    eprint = "2005.12299",
    archivePrefix = "arXiv",
    primaryClass = "hep-th",
    doi = "10.1103/RevModPhys.93.035003",
    journal = "Rev. Mod. Phys.",
    volume = "93",
    pages = "035003",
    year = "2021"
}

@article{Coquet:2023wjk,
    author = "Coquet, M. and Winn, M. and Du, X. and Ollitrault, J. Y. and Schlichting, S.",
    title = "{Dilepton Polarization as a Signature of Plasma Anisotropy}",
    eprint = "2309.00555",
    archivePrefix = "arXiv",
    primaryClass = "nucl-th",
    doi = "10.1103/PhysRevLett.132.232301",
    journal = "Phys. Rev. Lett.",
    volume = "132",
    pages = "232301",
    year = "2024"
}

@article{Bjorken:1982qr,
    author = "Bjorken, J. D.",
    title = "{Highly Relativistic Nucleus-Nucleus Collisions: The Central Rapidity Region}",
    doi = "10.1103/PhysRevD.27.140",
    journal = "Phys. Rev. D",
    volume = "27",
    pages = "140--151",
    year = "1983"
}

@article{Cacciari:1998it,
    author = "Cacciari, M. and Greco, M. and Nason, P.",
    title = "{The $p_T$ spectrum in heavy-flavour hadroproduction}",
    eprint = "hep-ph/9803400",
    archivePrefix = "arXiv",
    primaryClass = "hep-ph",
    doi = "10.1088/1126-6708/1998/05/007",
    journal = "JHEP",
    volume = "05",
    pages = "007",
    year = "1998"
}

@article{Catani:2020kkl,
    author = "Catani, S. and Devoto, S. and Grazzini, M. and Kallweit, S. and Mazzitelli, J.",
    title = "{Bottom-quark production at hadron colliders: fully differential predictions in NNLO QCD}",
    eprint = "2010.11906",
    archivePrefix = "arXiv",
    primaryClass = "hep-ph",
    doi = "10.1007/JHEP03(2021)029",
    journal = "JHEP",
    volume = "03",
    pages = "029",
    year = "2021"
}

@article{Klasen:2023uqj,
    author = "Klasen, M. and Paukkunen, H.",
    title = "{Nuclear PDFs After the First Decade of LHC Data}",
    eprint = "2311.00450",
    archivePrefix = "arXiv",
    primaryClass = "hep-ph",
    doi = "10.1146/annurev-nucl-102122-022747",
    journal = "Ann. Rev. Nucl. Part. Sci.",
    volume = "74",
    pages = "49--87",
    year = "2024"
}

@article{Kniehl:2012ti,
    author = "Kniehl, B. A. and Kramer, G. and Schienbein, I. and Spiesberger, H.",
    title = "{Inclusive Charmed-Meson Production at the CERN LHC}",
    eprint = "1202.0439",
    archivePrefix = "arXiv",
    primaryClass = "hep-ph",
    doi = "10.1140/epjc/s10052-012-2082-2",
    journal = "Eur. Phys. J. C",
    volume = "72",
    pages = "2082",
    year = "2012"
}

@article{Ethier:2020way,
    author = "Ethier, J. J. and Nocera, E. R.",
    title = "{Parton Distributions in Nucleons and Nuclei}",
    eprint = "2001.07722",
    archivePrefix = "arXiv",
    primaryClass = "hep-ph",
    doi = "10.1146/annurev-nucl-011720-042725",
    journal = "Ann. Rev. Nucl. Part. Sci.",
    volume = "70",
    pages = "43--76",
    year = "2020"
}

@article{Alwall:2014hca,
    author = "Alwall, J. and Frederix, R. and Frixione, S. and Hirschi, V. and Maltoni, F. and Mattelaer, O. and Shao, H. S. and Stelzer, T. and Torrielli, P. and Zaro, M.",
    title = "{The automated computation of tree-level and next-to-leading order differential cross sections, and their matching to parton shower simulations}",
    eprint = "1405.0301",
    archivePrefix = "arXiv",
    primaryClass = "hep-ph",
    doi = "10.1007/JHEP07(2014)079",
    journal = "JHEP",
    volume = "07",
    pages = "079",
    year = "2014"
}

@article{Miller:2007ri,
    author = "Miller, M. L. and Reygers, K. and Sanders, S. J. and Steinberg, P.",
    title = "{Glauber modeling in high energy nuclear collisions}",
    eprint = "nucl-ex/0701025",
    archivePrefix = "arXiv",
    primaryClass = "nucl-ex",
    doi = "10.1146/annurev.nucl.57.090506.123020",
    journal = "Ann. Rev. Nucl. Part. Sci.",
    volume = "57",
    pages = "205--243",
    year = "2007"
}

@article{Coquet:2021lca,
    author = "Coquet, M. and Du, X. and Ollitrault, J. Y. and Schlichting, S. and Winn, M.",
    title = "{Intermediate mass dileptons as pre-equilibrium probes in heavy ion collisions}",
    eprint = "2104.07622",
    archivePrefix = "arXiv",
    primaryClass = "nucl-th",
    doi = "10.1016/j.physletb.2021.136626",
    journal = "Phys. Lett. B",
    volume = "821",
    pages = "136626",
    year = "2021"
}

@article{ALICE:2015sru,
    author = "Adam, Jaroslav and others",
    collaboration = "ALICE",
    title = "{Rapidity and transverse-momentum dependence of the inclusive J/$\psi$ nuclear modification factor in p-Pb collisions at $ \sqrt{s_{N\ N}} =$ 5.02 TeV}",
    eprint = "1503.07179",
    archivePrefix = "arXiv",
    primaryClass = "nucl-ex",
    reportNumber = "CERN-PH-EP-2015-030",
    doi = "10.1007/JHEP06(2015)055",
    journal = "JHEP",
    volume = "06",
    pages = "055",
    year = "2015"
}

@article{Hou:2019efy,
    author = "Hou, Tie-Jiun and others",
    title = "{New CTEQ global analysis of quantum chromodynamics with high-precision data from the LHC}",
    eprint = "1912.10053",
    archivePrefix = "arXiv",
    primaryClass = "hep-ph",
    reportNumber = "MSUHEP-19-025, PITT-PACC-1911, SMU-HEP-19-03",
    doi = "10.1103/PhysRevD.103.014013",
    journal = "Phys. Rev. D",
    volume = "103",
    number = "1",
    pages = "014013",
    year = "2021"
}

@article{LHCb:2018roe,
    author = "Aaij, R. and others",
    title = "{Physics case for an LHCb Upgrade II - Opportunities in flavour physics, and beyond, in the HL-LHC era}",
    eprint = "1808.08865",
    archivePrefix = "arXiv",
    primaryClass = "hep-ex",
    year = "2018"
}

@article{dEnterria:2026tuz,
    author = "d'Enterria, David and Hekhorn, Felix and Helenius, Ilkka and Le, Van Dung and Paukkunen, Hannu",
    title = "{Inclusive charm and bottom quark pair production cross sections at hadron colliders at next-to-next-to-leading-order accuracy}",
    eprint = "2605.16019",
    archivePrefix = "arXiv",
    primaryClass = "hep-ph",
    month = "5",
    year = "2026"
}

@article{Nason:1989zy,
    author = "Nason, P. and Dawson, S. and Ellis, R. Keith",
    title = "{The One Particle Inclusive Differential Cross-Section for Heavy Quark Production in Hadronic Collisions}",
    reportNumber = "ETH-PT-89-2, FERMILAB-PUB-89-091-T, BNL-42398",
    doi = "10.1016/0550-3213(89)90286-1",
    journal = "Nucl. Phys. B",
    volume = "327",
    pages = "49--92",
    year = "1989",
    note = "[Erratum: Nucl.Phys.B 335, 260--260 (1990)]"
}

@article{Bodeker:2012zm,
    author = "Bodeker, D. and Laine, M.",
    title = "{Sommerfeld effect in heavy quark chemical equilibration}",
    eprint = "1210.6153",
    archivePrefix = "arXiv",
    primaryClass = "hep-ph",
    reportNumber = "BI-TP-2012-43",
    doi = "10.1007/JHEP01(2013)037",
    journal = "JHEP",
    volume = "01",
    pages = "037",
    year = "2013"
}

@article{ALICE-PUBLIC-2025-005,
      collaboration = "ALICEcollaboration",
      title         = "{European Strategy for Particle Physics 2026: Input from
                       the ALICE Collaboration}",
      year          = "2025",
      url           = "https://cds.cern.ch/record/2928780",
}

@article{Vagnoni:2025vmf,
    author = "Vagnoni, Vincenzo",
    collaboration = "LHCb",
    title = "{Heavy ion physics at LHCb Upgrade II}",
    eprint = "2503.23093",
    archivePrefix = "arXiv",
    primaryClass = "hep-ex",
    reportNumber = "LHCb-PUB-2025-003, CERN-LHCb-PUB-2025-003, LHCb-PUB-2025-003",
    month = "3",
    year = "2025"
}

@article{Adamova:2019vkf,
    author = "Adamov\'a, D. and Aglieri Rinella, G. and Agnello, M. and Ahammed, Z. and Aleksandrov, D. and Alici, A. and Alkin, A. and Alt, T. and Altsybeev, I. and Andreou, D. and others",
    title = "{A next-generation LHC heavy-ion experiment}",
    eprint = "1902.01211",
    archivePrefix = "arXiv",
    primaryClass = "physics.ins-det",
    year = "2019"
}

@article{Coquet:2021gms,
    author = "Coquet, M. and Du, X. and Ollitrault, J. Y. and Schlichting, S. and Winn, M.",
    title = "{Transverse mass scaling of dilepton radiation off a quark-gluon plasma}",
    eprint = "2112.13876",
    archivePrefix = "arXiv",
    primaryClass = "nucl-th",
    doi = "10.1016/j.nuclphysa.2022.122579",
    journal = "Nucl. Phys. A",
    volume = "1030",
    pages = "122579",
    year = "2023"
}

@article{Lappi:2006fp,
    author = "Lappi, T. and McLerran, L.",
    title = "{Some features of the glasma}",
    eprint = "hep-ph/0602189",
    archivePrefix = "arXiv",
    primaryClass = "hep-ph",
    doi = "10.1016/j.nuclphysa.2006.04.001",
    journal = "Nucl. Phys. A",
    volume = "772",
    pages = "200--212",
    year = "2006"
}

@article{Giacalone:2019ldn,
    author = "Giacalone, G. and Mazeliauskas, A. and Schlichting, S.",
    title = "{Hydrodynamic attractors, initial state energy and particle production in relativistic nuclear collisions}",
    eprint = "1908.02866",
    archivePrefix = "arXiv",
    primaryClass = "hep-ph",
    doi = "10.1103/PhysRevLett.123.262301",
    journal = "Phys. Rev. Lett.",
    volume = "123",
    pages = "262301",
    year = "2019"
}

@article{Du:2020dvp,
    author = "Du, X. and Schlichting, S.",
    title = "{Equilibration of weakly coupled QCD plasmas}",
    eprint = "2012.09079",
    archivePrefix = "arXiv",
    primaryClass = "hep-ph",
    doi = "10.1103/PhysRevD.104.054011",
    journal = "Phys. Rev. D",
    volume = "104",
    pages = "054011",
    year = "2021"
}

@article{Du:2020zqg,
    author = "Du, X. and Schlichting, S.",
    title = "{Equilibration of the Quark-Gluon Plasma at Finite Net-Baryon Density in QCD Kinetic Theory}",
    eprint = "2012.09068",
    archivePrefix = "arXiv",
    primaryClass = "hep-ph",
    doi = "10.1103/PhysRevLett.127.122301",
    journal = "Phys. Rev. Lett.",
    volume = "127",
    pages = "122301",
    year = "2021"
}

@article{ALICE:2016fbt,
    author = "Adam, J. and others",
    title = "{Centrality dependence of the pseudorapidity density distribution for charged particles in Pb-Pb collisions at $\sqrt{s_{\rm NN}}=5.02$ TeV}",
    eprint = "1612.08966",
    archivePrefix = "arXiv",
    primaryClass = "nucl-ex",
    doi = "10.1016/j.physletb.2017.07.017",
    journal = "Phys. Lett. B",
    volume = "772",
    pages = "567--577",
    year = "2017"
}

@article{Garcia-Montero:2023lrd,
    author = "Garcia-Montero, O. and Mazeliauskas, A. and Plaschke, P. and Schlichting, S.",
    title = "{Pre-equilibrium photons from the early stages of heavy-ion collisions}",
    eprint = "2308.09747",
    archivePrefix = "arXiv",
    primaryClass = "hep-ph",
    doi = "10.1007/JHEP03(2024)053",
    journal = "JHEP",
    volume = "03",
    pages = "053",
    year = "2024"
}

@article{Hanus:2019fnc,
    author = "Hanus, Patrick and Mazeliauskas, Aleksas and Reygers, Klaus",
    title = "{Entropy production in pp and Pb-Pb collisions at energies available at the CERN Large Hadron Collider}",
    eprint = "1908.02792",
    archivePrefix = "arXiv",
    primaryClass = "hep-ph",
    doi = "10.1103/PhysRevC.100.064903",
    journal = "Phys. Rev. C",
    volume = "100",
    number = "6",
    pages = "064903",
    year = "2019"
}

@article{AMY2003,
    doi = {10.1088/1126-6708/2003/01/030},
    url = {https://doi.org/10.1088/1126-6708/2003/01/030},
    year = {2003},
    publisher = {},
    volume = {2003},
    number = {01},
    pages = {030},
    author = {Arnold, Peter B.  and Moore, Guy D. and Yaffe, Laurence G.},
    title = {Effective kinetic theory for high temperature gauge theories},
    journal = {Journal of High Energy Physics},
}

@article{Plaschke2025,
  title = {Scaling of pre-equilibrium dilepton production in QCD kinetic theory},
  author = {Garcia-Montero, Oscar and Plaschke, Philip and Schlichting, S\"oren},
  journal = {Phys. Rev. D},
  volume = {111},
  issue = {3},
  pages = {034036},
  numpages = {9},
  year = {2025},
  month = {Feb},
  publisher = {American Physical Society},
  doi = {10.1103/PhysRevD.111.034036},
  url = {https://link.aps.org/doi/10.1103/PhysRevD.111.034036}
}

@article{ALICE:2023sgl,
    author = "Acharya, Shreyasi and others",
    collaboration = "ALICE",
    title = "{Charm production and fragmentation fractions at midrapidity in pp collisions at $ \sqrt{\textrm{s}} $ = 13 TeV}",
    eprint = "2308.04877",
    archivePrefix = "arXiv",
    primaryClass = "hep-ex",
    reportNumber = "CERN-EP-2023-162",
    doi = "10.1007/JHEP12(2023)086",
    journal = "JHEP",
    volume = "12",
    pages = "086",
    year = "2023"
}

@article{Eskola:2021nhw,
    author = "Eskola, Kari J. and Paakkinen, Petja and Paukkunen, Hannu and Salgado, Carlos A.",
    title = "{EPPS21: a global QCD analysis of nuclear PDFs}",
    eprint = "2112.12462",
    archivePrefix = "arXiv",
    primaryClass = "hep-ph",
    doi = "10.1140/epjc/s10052-022-10359-0",
    journal = "Eur. Phys. J. C",
    volume = "82",
    number = "5",
    pages = "413",
    year = "2022"
}

@article{Loizides:2017ack,
    author = "Loizides, Constantin and Kamin, Jason and d'Enterria, David",
    title = "{Improved Monte Carlo Glauber predictions at present and future nuclear colliders}",
    eprint = "1710.07098",
    archivePrefix = "arXiv",
    primaryClass = "nucl-ex",
    doi = "10.1103/PhysRevC.97.054910",
    journal = "Phys. Rev. C",
    volume = "97",
    number = "5",
    pages = "054910",
    year = "2018",
    note = "[Erratum: Phys.Rev.C 99, 019901 (2019)]"
}

@article{ALICE:2016yta,
    author = "Adam, Jaroslav and others",
    collaboration = "ALICE",
    title = "{$D$-meson production in $p$-Pb collisions at $\sqrt{s_{\rm NN}}=$5.02 TeV and in pp collisions at $\sqrt{s}=$7 TeV}",
    eprint = "1605.07569",
    archivePrefix = "arXiv",
    primaryClass = "nucl-ex",
    reportNumber = "CERN-EP-2016-127",
    doi = "10.1103/PhysRevC.94.054908",
    journal = "Phys. Rev. C",
    volume = "94",
    number = "5",
    pages = "054908",
    year = "2016"
}

@article{LHCb:2017yua,
    author = "Aaij, Roel and others",
    collaboration = "LHCb",
    title = "{Study of prompt D$^{0}$ meson production in $p$Pb collisions at $ \sqrt{s_{\mathrm{NN}}}=5 $ TeV}",
    eprint = "1707.02750",
    archivePrefix = "arXiv",
    primaryClass = "hep-ex",
    reportNumber = "LHCB-PAPER-2017-015, LHCb-PAPER-2017-015, CERN-EP-2017-135",
    doi = "10.1007/JHEP10(2017)090",
    journal = "JHEP",
    volume = "10",
    pages = "090",
    year = "2017"
}

@article{Churchill:2020uvk,
    author = "Churchill, Jessica and Yan, Li and Jeon, Sangyong and Gale, Charles",
    title = "{Emission of electromagnetic radiation from the early stages of relativistic heavy-ion collisions}",
    eprint = "2008.02902",
    archivePrefix = "arXiv",
    primaryClass = "hep-ph",
    doi = "10.1103/PhysRevC.103.024904",
    journal = "Phys. Rev. C",
    volume = "103",
    number = "2",
    pages = "024904",
    year = "2021"
}

@article{Wu:2024pba,
    author = "Wu, Xiang-Yu and Du, Lipei and Gale, Charles and Jeon, Sangyong",
    title = "{Probing the equilibration of the QCD matter created in heavy-ion collisions with dileptons}",
    eprint = "2407.04156",
    archivePrefix = "arXiv",
    primaryClass = "nucl-th",
    doi = "10.1103/PhysRevC.110.054904",
    journal = "Phys. Rev. C",
    volume = "110",
    number = "5",
    pages = "054904",
    year = "2024"
}

\end{document}